\documentclass{IEEEtran}

\usepackage{amsmath}
\usepackage{amssymb}
\usepackage{graphicx}

\newcommand{\set}[1]{{\cal #1}}
\newcommand{\sx}{\set{X}}
\newcommand{\sz}{\set{Z}}
\newcommand{\sS}{\set{S}}
\newcommand{\up}{\underline{p}}
\newcommand{\defined}{\triangleq}

\newtheorem{definition}{Definition}
\newtheorem{lemma}{Lemma}

\newtheorem{theorem}{Theorem}

\begin{document}

\title{Hidden Markov Process:
A New Representation,  Entropy Rate and Estimation
Entropy}

\author{Mohammad Rezaeian, ˜\IEEEmembership{Member,˜IEEE}
\thanks{Mohammad Rezaeian is with the  Department of Electrical and Electronic Engineering, University of Melbourne, Victoria,
3010, Email: rezaeian@unimelb.edu.au. This work was supported in
part by the Defense Advanced Research
   Projects Agency of the US Department of Defense and was monitored
   by the Office of Naval Research under Contract
   No. N00014-04-C-0437.}}

\maketitle

\begin{abstract}
We consider a pair of  correlated processes
$\{Z_n\}_{n=-\infty}^\infty$ and $\{S_n\}_{n=-\infty}^\infty$,
where the former is observable and the later is hidden. The
uncertainty in the estimation of $Z_n$ upon its finite past
history $Z_0^{n-1}$ is $H(Z_n|Z_0^{n-1})$, and for estimation of
$S_n$ upon this observation is $H(S_n|Z_0^{n-1})$, which are both
sequences of $n$. The limits of these sequences (and their
existence) are of practical  interests. The first limit, if
exists, is the entropy rate. We call the second limit the
\emph{estimation entropy}. An example of a process jointly
correlated to another one is the hidden Markov process. It is the
memoryless observation of the Markov state process where state
transitions are independent of past observations. We consider a
new representation of hidden Markov process using iterated
function system. In this representation the state transitions are
deterministically related to the process. By this representation
we analyze the two dynamical entropies for this process, which
results in integral expressions for the limits. This analysis
shows that under mild conditions the limits exist and provides a
simple method for calculating the elements of the corresponding
sequences.
\end{abstract}

\begin{keywords}
entropy rate, hidden Markov process, iterated function system,
estimation entropy.
\end{keywords}

\section{Introduction}
A stochastic process which is a noisy observation of a Markov
process through a memoryless channel is called a hidden Markov
process (HMP). In many applications of stochastic signal
processing such as radar and speech processing, the output of the
information source can be considered as an HMP. The entropy rate
of HMP as the limit of compressibility of information source thus
have special interest in those applications. Moreover, in the
additive noise channels the noise process can be characterized as
a hidden Markov process and its entropy rate is the defining
factor in the capacity of channel. Finding the entropy rate of the
hidden Markov process is thereby motivated by both applications in
stochastic signal processing, source coding and channel capacity
computation.

The study of the entropy rate of HMP started in 1957 by Blackwell
\cite{blkwell} who obtained an integral expression for the entropy
rate. This expression is defined through a measure described by an
integral equation which is hard to extract from the equation in
any explicit way. Bounds on the entropy rate can be computed based
on the conditional entropies on sets of finite number of random
variables \cite{ThCo}. Recent approaches for calculating the
entropy rate are Monte Carlo simulation \cite{HG} and Lyapunov
exponent \cite{HG},\cite{JSS}. However these approaches yield
indeterministic and hard to evaluate expressions. Simple
expression for the entropy rate has been recently obtained for
special cases where the parameters of hidden Markov source
approach zero \cite{JSS},\cite{OrWeiss}.

The hidden Markov process is a process defined through its
stochastic relation to another process. The entropy rate of HMP
thus corresponds to this relation and the dynamic of the
underlying process. However this entropy rate only indicates  the
residual uncertainty in the symbol one step ahead of observation
of the process itself. It doesn't indicate our uncertainty about
the underlying process. In this paper we define estimation entropy
as a variation of entropy rate to indicate this uncertainty. In
general for a pair of correlated processes which one of them is
hidden and the other is observable we can define estimation
entropy as the long run per symbol uncertainty in the estimation
of the hidden process based on the past observation. Such an
entropy measure will be an important criterion for evaluating the
performance of an estimator. In this paper we jointly analyze the
entropy rate and estimation entropy for a hidden Markov process.
This analysis is based on a mathematical model, namely the
iterated function system \cite{WSlom}, which suits the dynamics of
the information state process of the HMP. This analysis results in
integral expressions for these two dynamical entropies. We also
derive a numerical method for iteratively calculating entropy rate
and estimation entropy for HMP.

 In this paper a discrete random variable is denoted by upper
case and its realization by lower case. A sequence of random
variables $X_0,X_1,X_2,...X_n$ is denoted by $X_0^n$, whereas
$X^n$ refers to $X_{-\infty}^n$. The probability $Pr(X=x)$ is
shown by $p(x)$ (similarly for conditional probabilities), whereas
$\underline{p}(X)$ represents a row vector as the distribution of
$X$, ie: the $k$-th element of the vector $\up(X)$ is $Pr(X=k)$.
For a random variable $X$ defined on a set $\sx$, we denote by
$\nabla_\sx$ the probability simplex in $\mathbb{R}^{|\sx|}$. A
specific elements of a vector or matrix is referred to by its
index in square brackets or as  a subscript. The $z$-th row of
matrix $A$ is represented by $A^{(z)}$. The entropy of a random
variable $X$ is denoted by $H(X)$ whereas $h:
\nabla_\sx\rightarrow R^+$ represents the entropy function over
$\nabla_\sx$, i.e: $h(\up(X))=H(X)$ for all possible random
variables $X$ on $\sx$. Our notation does not distinguish
differential entropies from ordinary entropies.

In the next section we define the iterated function system and
draw some results from \cite{WSlom}, as well as a new  result. In
section III we define the hidden Markov process by identifying the
key properties for the probability distributions on the
corresponding domain sets and show that such a process can be
represented by an iterated function system . In sections IV and V
we
 derive integral expressions for
entropy rate and estimation entropy followed by a  method for
calculating them.

\section{Iterated Function Systems}
Consider a system with a state in the space of $\Delta$, where the
state transitions depends deterministically to a correlated
process taking values in a set $\mathcal{I}_m=\{1,2,...,m\}$ and
stochastically depending on the state.  The mathematical model
representing such a system is  an \emph{iterated function system}
(IFS) which is defined by $m$ functions transforming a metric
space $\Delta$ to itself, and $m$ place dependent probabilities.

\begin{definition}\label{def1}
A triple $\mathcal{F}=(\Delta,F_i,q_i)_{i=1,2...,K}$  is an
iterated function system if $F_i:\Delta\rightarrow \Delta$ and
$q_i:\Delta\rightarrow \mathcal{R}^+$ are measurable functions and
$\sum_iq_i=1$.
\end{definition}

The IFS represents the above mentioned dynamical system where the
probability of event $i\in \mathcal{I}_m$ under state $x\in\Delta$
is $q_i(x)$ and the consequence of such event is the change of
state to $F_i(x)$.

Although the generality of IFS allows the functions of $F_i$ and
$q_i$ to be measurable which is a wide range of real functions, in
this paper we are only interested in a subset of those functions,
the continuous functions. Such systems are referred to as
continuous IFS. If the functions $F_i$'s are only defined on
$\Delta_i\subset \Delta$, where
$\Delta_i=\{x\in\Delta:q_i(x)>0\}$, then the IFS is called partial
iterated function system (PIFS). Although the general application
of IFS  in this paper could be involved PIFS, we avoid such
complexity by restricting the application.

Consider $\mathcal{M}^1(\Delta)$ as the space of probability
measures on $\Delta$. For an $\mathcal{F}$ we define an operator
$\Phi:\mathcal{M}^1(\Delta)\rightarrow\mathcal{M}^1(\Delta),$
\begin{equation}\label{phidef}
  \Phi(\mu)(B)=\sum_i\int1_{B}(F_i(x))q_i(x)\mu(dx),
\end{equation}
for $\mu\in M^1(\Delta)$ and $B\subset \Delta$. The operator
$\Phi$, induced by $\mathcal{F}$,  represents the evolution of
probability measures under the action of $\mathcal{F}$. More
specifically, if our belief on the state of system at time $n$ is
the probability measure $\mu_n$,
($\mu_n\in\mathcal{M}^1(\Delta)$), then this belief at time $n+1$
is
\begin{equation}\label{mnn1}
  \mu_{n+1}=\Phi\mu_n,
\end{equation}
which can be easily verified by Equation \eqref{phidef} and role
of functions $F_i$ and $q_i$. Note that the operator $\Phi$ is
deterministic and it is affine, i.e:
$\Phi(\lambda\mu_1+(1-\lambda)\mu_2)=\lambda\Phi\mu_1+(1-\lambda)\Phi\mu_2$.
By such representation $\Phi$ is a so called \emph{Markov
operator}.

For a Markov operator $\Phi$ acting on the space
$\mathcal{M}^1(\Delta)$ a measure $\mu\in\mathcal{M}^1(\Delta)$ is
\emph{invariant} if $\mu=\Phi\mu$, and it is \emph{attractive} if
\begin{equation}\label{attractive}
\mu=\lim_{n\rightarrow\infty}\Phi^n\nu,
\end{equation}
for any $\nu\in\mathcal{M}^1(\Delta)$. A Markov operator $\Phi$
(and the corresponding IFS) is called \emph{asymptotically stable}
if it admits an invariant and attractive measure. The concept of
limit in Equations \eqref{attractive} is convergence in weak
topology, meaning
\begin{equation}\label{weaktop}
\int fd\mu=\lim_{n\rightarrow\infty}\int fd(\Phi^n\nu),
\end{equation}
for any continuous bounded function $f$. Note that the limit
doesn't necessarily exist or it is not necessarily unique. The set
of all attractive measures of $\Phi$ for $\mathcal{F}$ is denoted
by $\mathcal{S}^{\mathcal{F}}$.

A Markov operator which is continuous in weak topology is a Feller
operator. We can show that for a continuous IFS the operator
$\Phi$ is a Feller operator. In this case any
$\mu\in\mathcal{S}^{\mathcal{F}}$ is invariant.

Let $B(\Delta)$ be the space of all real valued continuous bounded
functions on $\Delta$. A special property of a Feller operator
$\Phi:\mathcal{M}^1(\Delta)\rightarrow\mathcal{M}^1(\Delta)$ is
that there exists an operator $\mathcal{U}:B(\Delta)\rightarrow
B(\Delta)$ such that:
\begin{equation}\label{intconj}
\int f(x)\Phi\mu(dx)=\int \mathcal{U}f(x)\mu(dx),
\end{equation}
for all $f\in B(\Delta), \mu\in\mathcal{M}^1(\Delta)$. The
operator $\mathcal{U}$ is called the operator conjugate to $\Phi$.
It can be shown \cite{WSlom} that for a continuous IFS the
operator conjugate of $\Phi$ is $\mathcal{U}$, where
\begin{equation}\label{opconj}
(\mathcal{U}f)(x)=\sum_{i\in \mathcal{I}_k} q_i(x)f(F_i(x)).
\end{equation}

For an IFS, the concept of change of state and probability of the
correlated process in each step can be extended to $n>1$ steps.
For an $\mathbf{i}=(i_1,i_2,...i_n)\in \mathcal{I}_m^n$, we denote
$$F_{\mathbf{i}}(x)=F_{i_n}(F_{i_{n-1}}(...F_{i_1}x)...))$$
$$q_{\mathbf{i}}(x)=q_{i1}(x)q_{i2}(F_{i1}(x))...q_{in}(F_{i_{n-1}}(F_{i_{n-2}}(...F_{i1}(x))))$$
Then the probability of the sequential event $\mathbf{i}$ under
state $x\in\Delta$ is $q_{\mathbf{i}}(x)$ and as a result of such
sequence, the state changes from $x$ to $F_{\mathbf{i}}(x)$ in $n$
steps. As an extension of \eqref{opconj}, we can show
\begin{equation}\label{opconjn}
(\mathcal{U}^nf)(x)=\sum_{\mathbf{i}\in \mathcal{I}_m^n}
q_{\mathbf{i}}(x)f(F_{\mathbf{i}}(x)).
\end{equation}

In this paper we define for a given continuous IFS, and for a
$f\in B(\Delta)$,
\begin{equation}\label{fhat}
 \hat{F}(x)\defined\lim_{n\rightarrow\infty}(\mathcal{U}^nf)(x).
\end{equation}

Now we state our result on IFS in the following Lemma which will
be used in Section IV as the  major application of IFS to the
purpose of this paper.

\begin{lemma}\label{IFSLem}
For a continuous IFS $\mathcal{F}=(\Delta,F_i,q_i)_{i=1,2...,K}$,
and any function $f\in B(\Delta)$,
\begin{equation}\label{lem1eq}
 \hat{F}(x)=\int fd\mu,
\end{equation}
where $\mu=\lim_{n\rightarrow\infty}\Phi^n\delta_x$ (if the limit
exists), and $\delta_x\in\mathcal{M}^1(\Delta)$ is a distribution
with all probability mass at $x$.
\end{lemma}
\begin{proof}
From \eqref{intconj} we have
$$\int f d(\Phi^2\mu)=\int\mathcal{U}fd(\Phi\mu)=\int(\mathcal{U}^2f)d\mu,$$
where the first equality is by substituting $\mu$ with $\Phi\mu$
in \eqref{intconj} and the second equality by substituting $f$
with $\mathcal{U}f$. Therefore by repetition of \eqref{intconj},
we have
\begin{equation}\label{rept2}
\int fd(\Phi^n\mu)=\int (\mathcal{U}^nf)d\mu,
\end{equation}
for all $f\in B(\Delta),\mu\in\mathcal{M}^1(\Delta)$. This results
in
$$\hat{F}(x)=\lim_{n\rightarrow\infty}\int(\mathcal{U}^nf)d\delta_x=
\lim_{n\rightarrow\infty}\int fd(\Phi^n\delta_x)=\int fd\mu,$$
where the first equality is from the definition of $\hat{F}$ in
\eqref{fhat} and the last one is from \eqref{weaktop}.

\end{proof}

From the above Lemma we infer that for an asymptotically stable
continuous IFS, the function $\hat{F}$ is a constant independent
of $x$. Note that  asymptotic stability ensures that there exists
at least one $\mu$ satisfying \eqref{weaktop} for any
$\nu\in\mathcal{M}^1(\Delta)$, which is true for $\nu=\delta_x$
for any $x$. If there are more than one
$\mu\in\mathcal{S}^\mathcal{F}$, all of them has to satisfy
\eqref{weaktop}. So in this case the Equality of \eqref{lem1eq}
independent of $x$ is true for any
$\mu\in\mathcal{S}^\mathcal{F}$.

We use the result of this section in the analysis of entropy
measures of hidden Markov processes by specializing $\Delta$ to be
the space of information state process and $f$ to be variations of
the entropy function.

\section{The Hidden Markov Process}

A hidden Markov process is a process related to an underlying
Markov process through a discrete memoryless channel, so it is
defined (for finite alphabet cases) by the transition probability
matrix $P$ of the Markov process and the emission matrix $T$ of
the memoryless channel \cite{Eph},\cite{Law}. In this paper the
hidden Markov process is referred to by
$\{Z_n\}_{n=-\infty}^\infty$, $Z_n\in\sz$ and its underlying
Markov process by $\{S_n\}_{n=-\infty}^\infty$, $S_n\in\sS$.  The
elements of matrices $P_{|\sS|\times|\sS|}$ and
$T_{|\sS|\times|\sz|}$ are the conditional probabilities,
\begin{equation}\label{PdefT}
  \begin{array}{rl}
P[s,s']&=p(S_{n+1}=s'|S_n=s),\\
T[s,z]&=p(Z_n=z|S_n=s) .
  \end{array}
\end{equation}

A pair of matrices $P$ and $T$  define a time invariant (but not
necessarily stationary) hidden Markov process on the state set
$\sS$ and observation set $\sz$ by the following basic properties,
for any $n$.
\begin{itemize}
    \item A1:\emph{ Markovity,}
\begin{equation}\label{mnsddf}
 p(s_n|s^{n-1})=p_P(s_n|s_{n-1}),
\end{equation}
where $p_P(s_n|s_{n-1})=P[s_{n-1},s_n]$.

\item A2:\emph{ Sufficient Statistics of State,}
    \begin{equation}\label{rgfds}
    p(s_n|s_{n-1},z^{n-1})=p_{P}(s_n|s_{n-1}),
\end{equation}
where $p_{P}(.|.)$ is defined by $P$.

\item A3:\emph{ Memoryless Observation,}
\begin{equation}\label{qwrfwa}
   p(z^n|s^n)=\prod_i^n p_{{T}}(z_i|s_i),
\end{equation}
where $p_{T}(z|s)=T[s,z]$.
\end{itemize}
Property A3 implies:
\begin{equation}\label{khde}
    p(z_n|s_n,z^{n-1})=p_T(z_n|s_n).
\end{equation}

For a hidden Markov process we define two random vectors $\pi_n$
and $\rho_n$ as functions of $Z^{n-1}$ on the domains $\nabla_\sS,
\nabla_\sz$, respectively,
\begin{equation}\label{ghdf}
    \pi_n(Z^{n-1})=\up(S_n|Z^{n-1}).
\end{equation}
\begin{equation}\label{resfd}
\rho_n(Z^{n-1})=\up(Z_n|Z^{n-1}).
\end{equation}
According to our notation,  the random vector $\pi_n$ has elements
$\pi_n[k]$, $k=1,2,...,|\sS|$,
\begin{equation*}\label{wer}
   \pi_n[k]=p(S_n=k|Z^{n-1}),
\end{equation*}
and similarly for $\rho_n$. We obtain the relation between random
vectors $\rho_n$ and $\pi_n$
\begin{equation}\label{matrp2}
\begin{array}{rl}
&\rho_n[m](Z^{n-1})\\
&=Pr(Z_n=m|Z^{n-1})\\
&=\sum_kPr(Z_n=m|Z^{n-1},S_n=k)Pr(S_n=k|Z^{n-1})\\
&=\sum_kPr(Z_n=m|S_n=k)Pr(S_n=k|Z^{n-1})\\
&=\sum_k T[k,m]\pi_n[k](Z^{n-1}),
\end{array}\end{equation}
 which shows the matrix relation
\begin{equation}\label{ghldf}
\rho_n=\pi_nT.
\end{equation}
More generally, we refer to $\zeta(\pi)\in\nabla_\sz$ as the
projection of $\pi\in\nabla_\sS$ under the mapping $T: \nabla_\sS
\rightarrow \nabla_\sz$, i.e:
\begin{equation}\label{ghdfdf}
\zeta(\pi)=\pi T.
\end{equation}

We can write
\begin{equation}\label{sufstat0}
    \up(Z_n|\pi_n,Z^{n-1})=\up(Z_n|Z^{n-1})=\rho_n=\zeta(\pi_n),
\end{equation}
where the first equality is due to $\pi_n$ being a function of
$Z^{n-1}$. Since the right hand side of \eqref{sufstat0} is (only)
a function of $\pi_n$ (and it is a distribution on $\sz$), the
left hand side must be equal to $\up(Z_n|\pi_n)$, i.e: we have
shown
\begin{equation}\label{sufstat}
    \up(Z_n|\pi_n)=\up(Z_n|\pi_n,Z^{n-1})=\zeta(\pi_n).
\end{equation}
This shows that $\pi_n$ is a sufficient statistics for the
observation process at time $n$. By a similar argument we have,
\begin{equation}\label{sufstat1}
    \up(S_n|\pi_n,Z^{n-1})=\up(S_n|Z^{n-1})=\pi_n=\up(S_n|\pi_n),
\end{equation}
which shows that $\pi_n$ is a sufficient statistics for the state
process at time $n$. In other words the random vector $\pi_n$
encapsulates all information about state at time $n$ that can be
obtained form all the past observations $Z^{n-1}$. For this reason
we call $\pi_n$ the \emph{information-state} at time $n$. A
similar definition for the information state with the same
property has been given  for the more general model of partially
observed Markov decision processes in \cite{SmaW}.

Using Bayes' rule and the law of total probability, an iterative
formula for the information state can be obtained as a function of
$z_n$, \cite{SmaW}, \cite{Gold},
\begin{equation}\label{iterpi}
  \pi_{n+1}=\eta(z_n,\pi_n),
\end{equation} where
\begin{equation}\label{uity}
    \eta(z,\pi)\defined \frac{\pi D(z)P}{\pi D(z)\underline{1}},
\end{equation}
where $D(z)$ is a diagonal matrix with $d_{k,k}(z)=T[k,z]$,
$k=1,2,..,|\sS|.$

Due to the sufficient statistic property of the information state,
we can consider the information state process
$\{\pi_n\}_{n=0}^\infty$ on $\nabla_\sS$ as the state process of
an iterated function system on $\nabla_\sS$ with the hidden Markov
process being its correlated process. This is because the hidden
Markov process at time $k$ is stochastically related to the
information state process at that time by
$Pr(Z_k=z|\pi_k=x)=\zeta(x)[z]$ (from \eqref {sufstat}). On the
other hand, $Z_k=z$ result in the deterministic change of state
from $\pi_k=x$ to $\pi_{k+1}=\eta(z,x)$. Consequently, for a
hidden Markov process there is a continuous iterated function
systems defined by, for different values $z\in\sz$,
\begin{equation}\label{ISfHMP}
 \begin{array}{rl}
F_z(x)&=\eta(z,x),\\
q_z(x)&=\zeta(x)[z],
 \end{array}
\end{equation}
where the equality $\sum_z q_z(x)=1, x\in\nabla_\sS$ is satisfied
due to $\zeta(x)\in\nabla_\sz$. These functions are in fact
conditional probabilities, $F_z(x)=\up(S_{k+1}|Z_k=z,\pi_{k}=x)$
and $q_z(x)=Pr(Z_k=z|\pi_k=x)$ for any $k$.

If the emission matrix $T$ has zero entries, then function
$\eta(z,x)$ could be indefinite for some $(z,x)$. This happens for
those $x\in\nabla_\sS$ that the element $z$ of vector $xT$ is
zero\footnote{e.g: if $T_{1,1}=T_{2,1}=0$, then for all $\pi$ that
have zero components on the third elements onward, both the
nominator and denominators of \eqref{uity} for $z=1$ will be zero,
and for those $\pi$'s the first component of $\pi T$ is zero.},
i.e: the functions $F_z(x)$ is only defined for $x$ that
$q_z(x)>0$. Hence for the general choice of matrix $T$ we have a
PIFS associated to the hidden Markov process. For this and other
reason that will reveals later we assume that matrix $T$ has non
zero entries.

For the continuous IFS related to the hidden Markov process, we
can obtain the corresponding Feller operator $\Phi$ and its
conjugate operator $\mathcal{U}$. The operator $\mathcal{U}$ maps
any $f\in B(\Delta)$ to $\mathcal{U}f\in B(\Delta)$ where
\begin{equation}\label{operu}
  \begin{array}{rl}
  (\mathcal{U}f)(x)&=\sum_z q_z(x)f(F_z(x)) \\
     =&\hspace{-.1in}\sum_z Pr(Z_k=z|\pi_k=x)f(\up(S_{k+1}|Z_k=z,\pi_{k}=x)).\
  \end{array}
\end{equation}

In general given $\pi_k=x$, the probability of a specific
$n$-sequence $\mathbf{z}=(z_1,z_2,...,z_n)$ for the HMP is
\begin{equation}\label{seqp}
  \begin{array}{l}
Pr(Z_{k}^{k+n-1}=\mathbf{z}|\pi_{k}=x)=\\
q_{z_1}(x)q_{z_2}(F_{z_1}(x))...
q_{z_{n}}(F_{z_{n-1}}(F_{z_{n-2}}(...F_{ 1}(x)))),
  \end{array}
\end{equation}
and this sequence changes the state
to
$$\pi_{k+n}=F_{z_n}(F_{z_{n-1}}(...F_{z_1}x)...)).$$
Therefore we can write
\begin{equation}\label{defseq}
\up(S_{k+n}|Z_{k}^{k+n-1}=\mathbf{z},\pi_{k}=x)=F_{z_n}(F_{z_{n-1}}(...F_{z_1}x)...)).
\end{equation}

Comparing to \eqref{opconjn}, we infer for any $f\in B(\Delta)$
and $\forall k$,
\begin{equation}\label{unf}
\begin{array}{l}
(\mathcal{U}^nf)(x)=\\
\sum_{\mathbf{z}}Pr(Z_{k}^{k+n-1}=\mathbf{z}|\pi_{k}=x)
f(\up(S_{k+n}|Z_{k}^{k+n-1}=\mathbf{z},\pi_{k}=x),
\end{array}
\end{equation}
For example, for entropy function $h$,
\begin{equation}\label{ggikg}
  h(x)\defined\sum_{i=1}^{|\sS|}-x[i]\log(x[i]),
\hspace{.2in}x\in\nabla_\sS,
\end{equation}we  have for any $ k$,
$$(\mathcal{U}^nh)(x)=H(S_{k+n}|Z_{k}^{k+n-1},\pi_{k}=x).$$
The IFS corresponding to a HMP under a wide range of the
parameters of the process is shown to be asymptotically stable.
\begin{definition}
A stochastic Matrix $P$ is primitive if there exists an $n$ such
that $(P^n)_{i,j}>0$ for all $i,j$.
\end{definition}

\begin{lemma}\label{HMPLem}
For a primitive matrix $P$ and an emission matrix $T$ with
strictly positive entries, the IFS $ $ defined according to
\eqref{ISfHMP} is asymptotically stable.
\end{lemma}
\begin{proof}
The proof follows from \cite[Theorem 8.1]{WSlom}. The IFS
$\mathcal{F}^P$ defined in \cite[Theorem 8.1]{WSlom} by
$$(F_i^P(x))_j\defined\frac{\sum_{l=1}^dx_lP_{lj}T_{li}}{\sum_{l=1}^dx_lT_{li}}=\eta(i,x)[j],$$
$$q_i^P(x)\defined\sum_{l=1}^dx_lT_{li}=\zeta(x)[i],$$
is the same as the IFS defined by \eqref{ISfHMP}. It is shown in
\cite[Theorem 8.1]{WSlom} that under the conditions of this lemma
$\mathcal{F}^P$ is asymptotically hyperbolic, which then has to be
asymptotically stable according to \cite[Theorem 3.4]{WSlom}.
\end{proof}
A Markov chain with primitive transition matrix $P$ is
geometrically ergodic and has a unique stationary distribution
\cite{Eph}.

\section{Entropy Rate and Estimation Entropy}
The entropy of a random variable $Z\in\sz$ is a function of its
distribution $\up(Z)\in\nabla_Z$,
\begin{equation*}
  \begin{array}{rl}
H(Z)=h(\up(Z))=\sum_z-p(z)\log p(z).
  \end{array}
\end{equation*}
For a general process $\{Z_n\}_{n=-\infty}^{\infty}$, the entropy
of any $n$-sequence $Z_{k}^{k+n-1}$ is denoted by
$H(Z_{k}^{k+n-1})$ which is defined by the joint probabilities
$Pr(Z_{k}^{k+n-1}=\mathbf{z})$, for all $\mathbf{z}\in\sz^n$. For
a stationary process these joint probabilities are invariant with
$k$. The entropy rate of the
 process is denoted by $\hat{H}_Z$ and defined as
\begin{equation}\label{entradef}
\hat{H}_Z\defined\lim_{n\rightarrow\infty}\frac{1}{n}H(Z_0^n),
\end{equation}when the limit exists.
Let $$\sigma_n\defined H(Z_n|Z_0^{n-1})=H(Z_0^{n})-H(Z_0^{n-1}).$$
We see that the entropy rate is the limit of Cesaro mean of the
sequence of $\sigma_n$, i.e:
\begin{equation}\label{cesaro}
\hat
H_Z=\lim\limits_{n\rightarrow\infty}\frac{1}{n}\sum_{i=1}^{n}\sigma_i.
\end{equation}
We know that if the sequence of $\sigma_n$ converges, then the
sequence of its Cesaro mean also converges to the same limit
\cite[Theorem 4.2.3]{ThCo}. However the opposite is not
necessarily true. Therefore, the entropy rate is equal to
\begin{equation}\label{entrat}
\hat{H}_Z=\lim_{n\rightarrow\infty}H(Z_n|Z_0^{n-1}),
\end{equation}
when this limit exists, but the non-existence of this limit
doesn't mean that the entropy rate doesn't exist. On the other
hand, the sequence of $\sigma_n$ converges faster than the
sequence in \eqref{cesaro} to its limit. Therefore the convergence
rate of \eqref{entrat} is faster than \eqref{entradef}. This fact
was first pointed out in \cite{Shannon}.

One sufficient condition for the existence of the limit of
$\sigma_n$ is the stationarity of the process. For a stationary
process
\begin{equation}\label{stationarity}
  \sigma_n=H(Z_{n+1}|Z_1^{n})\ge
H(Z_{n+1}|Z_0^{n-1})=\sigma_{n+1}\ge 0,
\end{equation}
which shows that $\sigma_n$ must have a limit. Therefore for a
stationary process we can write entropy rate as \eqref{entrat}.
For a stationary Markov process with transition matrix $P$ the
entropy rate is
\begin{equation}\label{MarkEnt}
\hat{H}_Z=\lim_{n\rightarrow\infty}H(Z_n|Z_{n-1})=H(Z_1|Z_0)=\sum_ix[i]h(P^{(i)}),
\end{equation}
where $x\in\nabla_\sz$ is the stationary distribution of the
Markov process, i.e: the solution of xP=x. Of special interest to
this paper is the entropy rate of the hidden Markov process.

We can extend the concept of entropy rate to a pair of correlated
processes. Assume we have a jointly correlated processes
$\{Z_n\}_{n=-\infty}^{\infty}$ and $\{S_n\}_{n=-\infty}^{\infty}$
where we observe the first process and based on our observation
estimate the state of the other process. The uncertainty in the
estimation of $S_n$ upon past observations $Z_0^{n-1}$ is
$H(S_n|Z_0^{n-1})$. The limit of this sequence which inversely
measures the observability of the hidden process is of practical
and theoretical interests.
 We call this limit \emph{Estimation
Entropy},
\begin{equation}\label{estent}
  \hat{H}_{S/Z}\defined\lim_{n\rightarrow\infty}H(S_n|Z_0^{n-1}),
\end{equation}
when the limit exists. Similar to entropy rate, we can consider
the limit of Cesaro mean of the sequence $\beta_n\defined
H(S_n|Z^{n-1})$ (i.e:
$\lim\limits_{n\rightarrow\infty}1/n\sum_{i=1}^n\beta_i$ ) as the
estimation entropy, which gives a more relaxed condition on its
existence, but it will have a much slower convergence rate.
However, if both limits exist, then they will be equal. If the two
processes $\{Z_n\}_{n=-\infty}^{\infty}$ and
$\{S_n\}_{n=-\infty}^{\infty}$ are jointly stationary, then
$\beta_n$ is decreasing and non-negative (same as
\eqref{stationarity}), thus the limit in \eqref{estent} exists. We
see that for a wide range of non-stationary processes also the
limits in \eqref{entrat} and \eqref{estent} exist.

Practical application of estimation entropy
 is for example in sensor scheduling for observation of a
Markov process \cite{Ray}. The aim of such a scheduler is to find
a policy for selection of sensors based on information-state which
minimizes the estimation entropy, thus achieving the maximum
observability for the Markov process. This entropy measure could
also be related to the error probability in channel coding. The
more the estimation entropy, the more uncertainty per symbol in
the decoding process of the received signal, thus higher error
probability.  The estimation entropy can be viewed as a benchmark
for indicating how well an estimator is working. It is the limit
of minimum uncertainty that an estimator can achieve for
estimating the current value of the unobserved process under the
knowledge of enough history of observations. We consider HMP as a
joint process and analyze its estimation entropy.

For a stationary hidden Markov process the entropy rate
$\hat{H}_Z$ and estimation entropy $\hat{H}_{S/Z}$ are the
limiting expectations
\begin{equation}\label{limexp}
\begin{array}{rl}
\hat{H}_Z&=\lim\limits_{n\rightarrow\infty}E[h(\rho_n)],\\
\hat{H}_{S/Z}&=\lim\limits_{n\rightarrow\infty}E[h(\pi_n)].
\end{array}
\end{equation}
However since $\pi_n$ and $\rho_n$ are functions of joint
distributions of random variables $Z_0^{n-1}$ these expectations
are not directly computable. We use the IFS for a hidden Markov
process to gain insight into  these entropy measures in a more
general setting without the stationarity assumption.

 Adapting Equation
\eqref{phidef} with special functions $F_z(x)$ and $q_z(x)$ in
\ref{ISfHMP}, we obtain the Feller operator $\Phi$ for the IFS
corresponding to a hidden Markov process.

\begin{equation}\label{phin}
\Phi(\mu)(B)=\sum_z\int\limits_{\nabla_\sS}1_B(\eta(z,x))\zeta(x)[z]\mu(dx).
\end{equation}

To analyze the entropy measures $\hat{H}_Z$ and $\hat{H}_{S/Z}$,
we define two intermediate functions
\begin{equation}\label{itermediates}
\begin{array}{rl}
  \hat{H}_Z(x)&=\lim\limits_{n\rightarrow\infty}H(Z_n|Z_0^{n-1},\pi_0=x),\\
\hat{H}_{S/Z}(x)&=\lim\limits_{n\rightarrow\infty}H(S_n|Z_0^{n-1},\pi_0=x).
\end{array}
\end{equation}
In comparison to \eqref{entrat} and \eqref{estent}, these
functions are the corresponding per symbol entropies when it is
conditioned on a specific prior distribution of state at time
$n=0$. We now use Lemma \ref{IFSLem} to obtain an integral
expressions for these limiting entropies.
\begin{lemma}\label{EntLem}
For a hidden Markov process
\begin{equation}\label{the1}
 \begin{array}{rl}
   \hat{H}_Z(x)&= \int\limits_{\nabla_\sS} (h_1\circ\zeta) d\mu,\\
   \hat{H}_{S/Z}(x) &= \int\limits_{\nabla_\sS}  h_2 d\mu, \\
 \end{array}
\end{equation}
where $\mu=\lim_{n\rightarrow\infty}\Phi^n\delta_x$, and
$h_1:\nabla_\sz\rightarrow\mathbb{R}^+$ and
$h_2:\nabla_\sS\rightarrow\mathbb{R}^+$ are entropy functions.
\end{lemma}
\begin{proof}From definition of conditional entropy we write,
\begin{equation}\label{urdf3fnf}
\begin{array}{l}
H(Z_n|Z_0^{n-1},\pi_0=x)=\\
\sum_{\mathbf{z}}Pr(Z_{0}^{n-1}=\mathbf{z}|\pi_{k}=x)
h_1(\up(Z_{n}|Z_{0}^{n-1}=\mathbf{z},\pi_{0}=x)).
\end{array}
\end{equation}
Now since (as in \eqref{matrp2}, using
$p(z_n|s_n,z^{n-1},\pi_0)=p(z_n|s_n)$),
\begin{equation}\label{trans}
  \up(Z_{n}|Z_{0}^{n-1}=\mathbf{z},\pi_{0}=x)=
\zeta(\up(S_{n}|Z_{0}^{n-1}=\mathbf{z},\pi_{0}=x)),
\end{equation}
 Equation
\eqref{urdf3fnf} can  be written as
\begin{equation}\label{urdf3fnf2}
\begin{array}{l}
H(Z_n|Z_0^{n-1},\pi_0=x)=\\
\sum_{\mathbf{z}}Pr(Z_{0}^{n-1}=\mathbf{z}|\pi_{k}=x)
h_1\circ\zeta(\up(S_{n}|Z_{0}^{n-1}=\mathbf{z},\pi_{0}=x)).
\end{array}
\end{equation}
Similarly from definition of conditional entropy, we can write
\begin{equation}\label{urdffnf}
\begin{array}{l}
H(S_n|Z_0^{n-1},\pi_0=x)=\\
\sum_{\mathbf{z}}Pr(Z_{0}^{n-1}=\mathbf{z}|\pi_{k}=x)
h_2(\up(S_{n}|Z_{0}^{n-1}=\mathbf{z},\pi_{0}=x)).
\end{array}
\end{equation}
Comparing Equations \eqref{urdf3fnf2} with \eqref{unf}, we have
\begin{equation}\label{1eqw}
\hat{H}_Z(x)=\lim\limits_{n\rightarrow\infty}(\mathcal{U}^n(h_1\circ\zeta))(x).
\end{equation}

Similarly by \eqref{urdffnf},
\begin{equation}\label{2eqw}
\hat{H}_{S/Z}(x)=\lim\limits_{n\rightarrow\infty}(\mathcal{U}^nh_2)(x).
\end{equation}
Now considering Equation \eqref{fhat} and applying Lemma
\ref{IFSLem} we obtain \eqref{the1}.
\end{proof}

Lemmas \ref{HMPLem} and \ref{EntLem} result in integral
expressions for entropy rate and estimation entropy.

\begin{theorem}\label{theorem1}
For a hidden Markov process with primitive matrix $P$ and the
emission matrix $T$ with strictly positive entries,
\begin{equation}\label{the21}
 \begin{array}{rl}
   \hat{H}_Z&= \int\limits_{\nabla_\sS} (h_1\circ\zeta) d\mu,\\
   \hat{H}_{S/Z} &= \int\limits_{\nabla_\sS}  h_2 d\mu, \\
 \end{array}
\end{equation}
where $\mu$ is any attractive and invariant measure of operator
$\Phi$, and $h_1,h_2$ are the entropy functions on
$\nabla_\sz,\nabla_\sS$, respectively.
\end{theorem}
\begin{proof}
From Lemma \ref{HMPLem}, under the condition of this Theorem, the
continuous IFS corresponding to the HMP is asymptotically stable.
As it is discussed after Lemma\ref{IFSLem}, in this case the
functions $\hat H_Z(x)$ and $\hat H_{S/Z}(x)$ (in \eqref{1eqw} and
\eqref{2eqw}) are independent of $x$ and the equalities of
\eqref{the1} are satisfied for any attractive measure (which
exists and it is also an invariant measure) of $\Phi$. The
independency of $x$ for $\hat{H}_Z(x)$ and $\hat{H}_{S/Z}(x)$ in
\eqref{itermediates} results in the equalities in \eqref{the21}
for $\hat{H}_Z$ and $\hat{H}_{S/Z}$. Note that for a set of random
variables $X,Y,Z$ if $H(Y|Z,X=x)$ is invariant with $x$, then
$H(Y|Z)=H(Y|Z,X)=H(Y|Z,X=x)$. Moreover from the existence of limit
of $\sigma_n$ (defined before) this limit is equal to $\hat H_Z$.
\end{proof}

The first equality in the above theorem has been previously
obtained by a different approach in \cite{reza05}. However in
\cite{reza05}, the measure $\mu$ is restricted to be
$\mu=\lim\limits_{n\rightarrow\infty} \Phi^n\delta_{x^*}$, where
$x^*$ is the stationary distribution of the underlying Markov
process defined by $P$,
\begin{equation}\label{xstar}
x^*P=x^*.
\end{equation}
The integral expression for $\hat H_Z$ in Theorem \ref{theorem1}
is also the same as the expression in \cite[Proposition
8.1]{WSlom} for $\theta=P$.  For this case the integral expression
is shown to be equal to both of the following two entropy measures
\begin{equation}\label{hcal}
\begin{array}{rl}
\mathcal{H}(x^*)&\defined\lim\limits_{n\rightarrow\infty}\frac{1}{n}
\sum_{\mathbf{z}\in\sz^n}
q_{\mathbf{z}}(x^*)\log(q_\mathbf{z}(x^*)),\\
\mathcal{H}(\mu)&\defined\lim\limits_{n\rightarrow\infty}\frac{1}{n}
\sum_{\mathbf{z}\in\sz^n} \int q_{\mathbf{z}}(x)\mu(dx).\log(\int
q_{\mathbf{z}}(x)\mu(dx)),
\end{array}
\end{equation}
where $\mu$ is the  attractive and invariant measure of $\Phi$ for
the IFS defined by \eqref{ISfHMP}. Considering
$q_{\mathbf{z}}(x)=p(Z_0^{n-1}=\mathbf{z}|\pi_0=x)$ for HMP, (c.f.
\eqref{seqp}), the two equalities match with  Lemma \ref{EntLem}
and Theorem \ref{theorem1}. However, the analysis in \cite{WSlom}
is based on a general and complex view to dynamical systems, where
the dynamics of system is represented by a Markov operator and the
measurement process is separately represented by a Markov pair,
and this Markov pair corresponds to a PIFS.

The integral expression for $\hat H_Z$ is also equivalent to the
original Blackwell's formulation \cite{blkwell} by a change of
variable $x$ to $xP$. This is because the expression in
\cite{blkwell} is derived based on
$\alpha_{n-1}=\up(S_{n-1}|Z^{n-1})$ instead of
$\pi_n=\alpha_{n-1}P$ in \eqref{ghdf} (cf. \eqref{rgfds}). The
measure of integral also corresponds to this change of variable.
Note that the measure $\mu$ in \eqref{the21} satisfies (due to its
invariant property)
\begin{equation}\label{blw}
  \mu(B)=\Phi(\mu)(B)=\sum_z\int_{F_z^{-1}(B)}(xT)[z]\mu(dx),
\end{equation}
(cf. \eqref{phin}) which is the same as the integral equation for
the measure in \cite{blkwell} if we change the integrand of
\eqref{blw} to $r_z(x)=(xPT)[z]$ and instead of $F_z(x)$ use the
function $f_z(x)=xPD(z)/r_z(x)$ (derived from \eqref{uity} by
$\pi=\alpha P$, satisfying $\alpha_{n+1}=f_z(\alpha_{n})$).

\section{A Numerical Algorithm}
Here we obtain a numerical method for computing entropy rate and
estimation entropy based on  Lemma \ref{EntLem} and the fact that
with the condition of Theorem \ref{theorem1}, \eqref{the1} is
independent of $x$. The computational complexity of this method
grows exponentially with the iterations,
 but numerical examples show a very fast convergence. In \cite{rezaISIT06}
 it is shown that applying this method for computation of entropy rate
 yields the same capacity results for  symmetric Markov
 channels similar to previous results.

We write \eqref{the1} as
\begin{equation}\label{lemdffnew}
 \begin{array}{rl}
   \hat{H}_Z(\nu)&= \lim\limits_{n\rightarrow\infty}\int\limits_{\nabla_\sS} (h_1\circ\zeta) d\mu_n,\\
   \hat{H}_{S/Z}(\nu) &= \lim\limits_{n\rightarrow\infty}\int\limits_{\nabla_\sS}  h_2 d\mu_n, \\
 \end{array}
\end{equation}
where $\mu_n=\Phi^n\delta_\nu$.
 Considering $\overline{\mu}_n:
\nabla_\sS\rightarrow\mathcal{R}$ as the probability density
function corresponding to the probability measure $\mu_n$, from
\eqref{mnn1} and \eqref{phin} we have the following recursive
formula
\begin{equation}\label{iu765sd}
    \overline{\mu}_{n+1}(\pi_{n+1})= \sum_z\int\limits_{\nabla_\sS}
\delta(\pi_{n+1}-\eta(z,\pi_n))\zeta(\pi_n)[z]\overline{\mu}_n(\pi_n)d\pi_n.
\end{equation}
Corresponding to the initial probability measure $\delta_{\nu}$,
we have the initial density function
$\overline{\mu}_0(x)=\delta(x-\nu)$. By $\overline{\mu}_0$ being a
probability mass function, Equation \eqref{iu765sd} yields a
probability mass function $\overline{\mu}_n$ for any $n$. For
example $\overline{\mu}_1(.)$ is
\begin{equation*}
\overline{\mu}_1(\pi_1)=\sum_z
\delta(\pi_1-\eta(z,\nu))\zeta(\nu)[z],
\end{equation*}
which is a $|\sz|$ point probability mass function. By induction
it can be shown that the distribution $\overline{\mu}_n(.)$ for
any $n$ is a probability mass function  over a finite set $U_{n}$
which consists of $|\sz|^n$ points of $\nabla_{\sS}$,
$U_{n}=\{u\in \nabla_{\sS}:u=\eta(z,v),z\in\sz,v\in U_{n-1}\}$,
$|U_n|=|\sz|^n$, $U_0=\{\nu\}$. The probability distribution over
$U_n$ is $\dot{\mu}_n(u)=\dot{\mu}_{n-1}(v)\zeta(v)[z]$ for
$u=\eta(z,v)$, $v\in U_{n-1}$. Therefore for every $v\in U_{n-1}$,
$|\sz|$ points will be generated in $U_n$ that corresponds to
$\eta(z,v)$ for different $z$, and the probability of each of
those points will be $\dot{\mu}_{n-1}(v)(vT)[z]$.

Starting from $U_0=\{\nu\}$ for some $\nu\in\nabla_\sS$, by the
above method we can iteratively generate the sets $U_n$ and the
probability distribution $\dot{\mu}_n(.)$ over these sets. The
integrals in \eqref{lemdffnew} can now be written as summation
over $U_n$, therefore the entropy rate and estimation entropy are
the limit of the following sequences
\begin{equation}\label{rtk60}
\begin{array}{rl}
 H_Z^n&=\sum_{i=1}^{|\sz|^n} \dot{\mu}_n(u_i)h_1(u_iT), \hspace{.2in}
  u_i\in U_n,\\
 H_{S/Z}^n&=\sum_{i=1}^{|\sz|^n} \dot{\mu}_n(u_i)h_2(u_i), \hspace{.2in}
  u_i\in U_n,
\end{array}
\end{equation}
where
\begin{equation*}
  \begin{array}{rl}
h_1(\rho)&=-\sum_z \rho[z]\log \rho[z]$, $\rho\in\nabla_\sz,\\
h_2(\pi)&=-\sum_s \pi[s]\log \pi[s]$, $\pi\in\nabla_\sS.
  \end{array}
\end{equation*}

\begin{figure}
\centering
\includegraphics[width=3.5in]{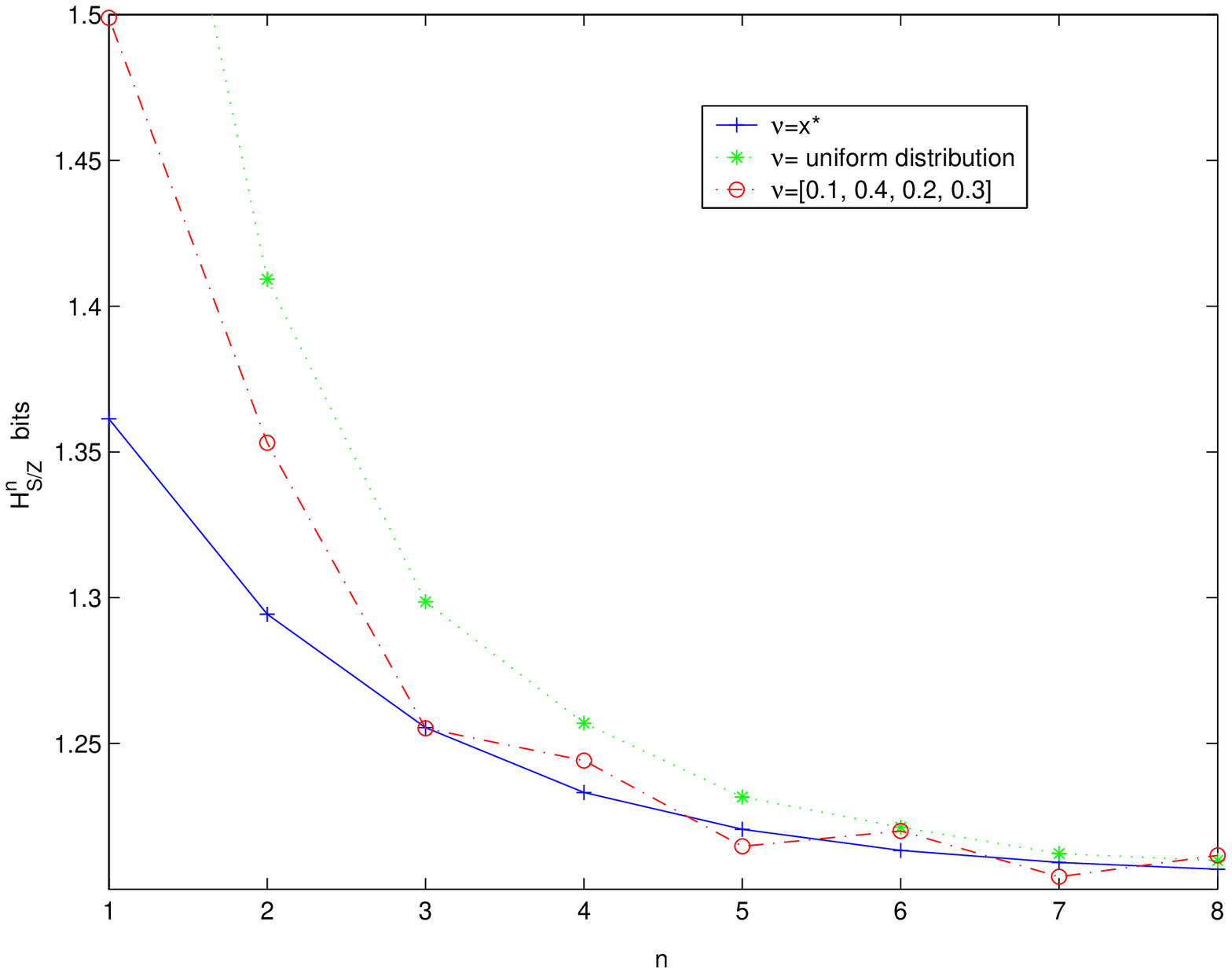} \\
  \caption{The convergence of the proposed algorithm to the entropy rate (left) and
  estimation entropy  of the example hidden Markov process for various $\nu$.}\label{convergence}
\label{Fig_estentropy}
\end{figure}

Figure \ref{Fig_estentropy} shows the convergence of the proposed
method to the entropy rate and estimation entropy for various
starting points $\nu$ for an example hidden Markov process. In
this example $\sS=\sz=\{0,1,2,3\}$, and
\begin{equation*}
P=
\left(%
\begin{array}{cccc}
.02& .03& .05 &.9\\
 .8& .06& .04 &.1\\
  .1 &.7 &.15 &.05\\
  .9& .03 &.03 &.04
\end{array}%
\right), T=
\left(%
\begin{array}{cccc}
.1 &.2 &.5 &.2\\
.6 &.1& .2& .1\\
 .5& .2& .1& .2\\
 .3 &.2 &.1 &.4
 \end{array}%
\right).
\end{equation*}
Although the result of Section IV ensures convergence of algorithm
for any starting distribution $\nu$, this figure and other
numerical examples show faster convergence for $\nu=x^*$ (the
solution of \eqref{xstar}). Without the condition of Theorem
\ref{theorem1}, the convergence could be to different values for
various $\nu$. Among various examples of HMP, the convergence will
be slower where the entropy rate of the underlying Markov process
with transition probability matrix $P$ ($\hat H_Z$ in
\eqref{MarkEnt}) is very low relative to $\log_2|\sS|$ (in the
above example it is 0.678b relative to 2b) or the rows of $T$ have
high entropy.

The sequence of $H^n_Z$, as the right hand side of
\eqref{lemdffnew} for finite $n>0$, is in fact
$H_Z^n=H(Z_n|Z_0^{n-1},\pi_0=\nu)$. If we assume (as in
\cite{ThCo}) that the process $Z_n$ starts at time zero, i.e: one
sided stationary process, then $\pi_0$ means the distribution of
state without any observation which if we further assume that it
is the stationary distribution of state process, i.e: $x^*$ in
\eqref{xstar}, then both of the processes
$\{Z_n\}_{n=-\infty}^\infty$ and $\{S_n\}_{n=-\infty}^\infty$ are
stationary. So for $\nu=x^*$, $H_Z^n=H(Z_n|Z_0^{n-1})=\sigma_n$,
and similarly $H_{S/Z}^n=H(S_n|Z_0^{n-1})=\beta_n$, and the
sequences of $\sigma_n$ and $\beta_n$ converge monotonically from
above to their limits. Therefore, $H_Z^n$ and $H_{S/Z}^n$ as
defined in \eqref{rtk60} for $\nu=x^*$ are always  monotonically
decreasing sequence of $n$. Figure \ref{Fig_estentropy}
exemplifies this fact.

\begin{figure} \centering
\includegraphics[width=3.5in]{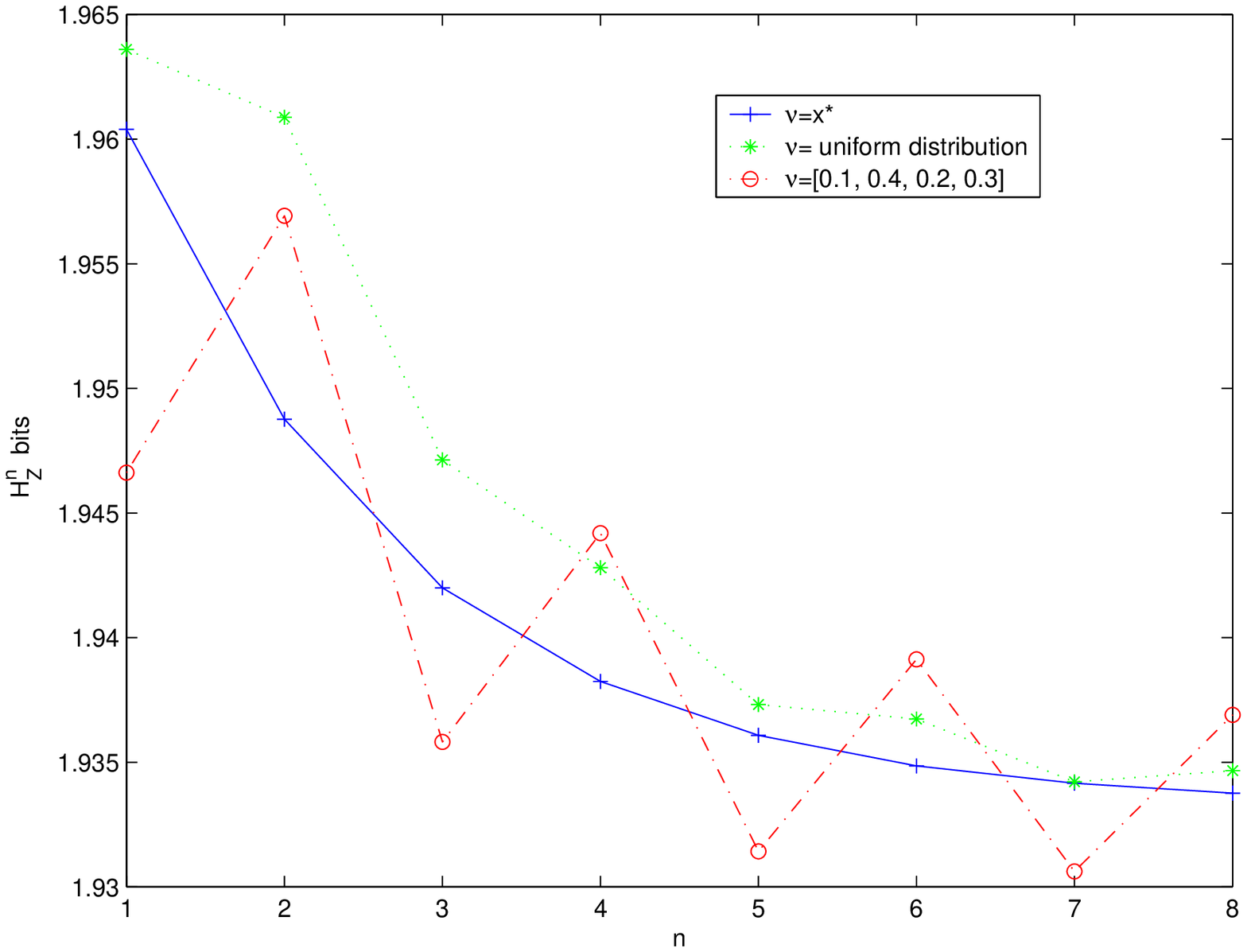} \\
\label{Fig_entropy}
\end{figure}
\section{Conclusion}
 HMP is a process described by its
relation to a Markov state process which has stochastic transition
to the next state independent of the current realization of the
process.  In this paper we showed that HMP can be better described
and more rigorously analyzed by iterated function systems whose
state transitions are deterministically related to the process. In
both descriptions the state is hidden and the process at any time
is stochastically related to the state at that time.

In this paper we also introduced the concept of estimation entropy
for a pair of joint processes which has practical applications.
The entropy rate for a process, like HMP, which is correlated to
another process can be viewed as the self estimation entropy. Both
entropy rate and estimation entropy for the hidden Markov process
can be  analyzed using the iterated function system description of
the process. This analysis results in integral expressions for
these dynamical entropies. The integral expressions are based on
an attractive and invariant measure of the Markov operator induced
by the iterated function system. These integrals can be evaluated
numerically as the limit of special numerical sequences.

\section{Acknowledgment}
The author would like to thank Wojciech Slomczynski for bringing
to attention the underpinning theories of this paper from his
eminent monograph \cite{WSlom}.  The special application of
estimation entropy to scheduling problem \cite{Ray} is a joint
work with Bill Moran and Sofia Suvorova.
\bibliographystyle{IEEEtran}

\end{document}